\PassOptionsToPackage{table}{xcolor}
\documentclass[a4paper,twocolumn,10pt]{article}

\usepackage{hegroup}
\usepackage{times}
\usepackage{latexsym}
\usepackage{multirow}
\usepackage{array}
\usepackage{makecell}
\usepackage{colortbl}
\usepackage{xspace}
\usepackage{algorithm}
\usepackage{tabularx}
\usepackage{tcolorbox}
\tcbuselibrary{listings}
\usepackage{algorithmic}
\usepackage{amsmath}
\usepackage{appendix}
\usepackage{bbm}
\usepackage{amssymb}
\usepackage{float}
\usepackage{cleveref}
\newcommand{\framework}{\textsc{CIPHER}\xspace}
\newcommand{\mypara}[1]{\noindent{\bf {#1}.}\xspace}
\usepackage{microtype}
\usepackage{inconsolata}
\usepackage{graphicx}
\usepackage{tikz}
\usepackage{pgfplots}
\pgfplotsset{compat=1.18}

\title{CIPHER: Cross-Paper Implementation and Pipeline Harmonization for Evaluation and Reproduction}
\date{}

\author{
Leyi Sheng\textsuperscript{1}\thanks{Equal contribution.}, \quad
Han Sun\textsuperscript{2,3}\thanksmark{1}, \quad
Zhen Sun\textsuperscript{1}, \\
Yuntao Yue\textsuperscript{1,5}, \quad
Jinlin Wu\textsuperscript{6,7}, \quad
Xinlei He\textsuperscript{4}\thanksmark{2}, \quad
Jiaheng Wei\textsuperscript{1}\thanks{Corresponding authors: Jiaheng Wei (\protect\href{mailto:jiahengwei@hkust-gz.edu.cn}{jiahengwei@hkust-gz.edu.cn}), Xinlei He (\protect\href{mailto:xinlei.he@whu.edu.cn}{xinlei.he@whu.edu.cn}).}
\\
\textsuperscript{1}\textit{The Hong Kong University of Technology and Science (Guangzhou)} \\
\textsuperscript{2}\textit{East China Normal University}, \quad
\textsuperscript{3}\textit{Shanghai Qi Zhi Institute}, \quad
\textsuperscript{4}\textit{Wuhan University} \\
\textsuperscript{5}\textit{Institute of Deep Perception Technology, JITRI} \\
\textsuperscript{6}\textit{CAIR, Hong Kong Institute of Science and Innovation (HKISI)} \\
\textsuperscript{7}\textit{MAIS, Institute of Automation, Chinese Academy of Sciences}
}

\begin{document}
\maketitle
\begin{abstract}
Reproducing text-to-image (T2I) jailbreak methods requires integrating diverse attack algorithms with image generation, safety filtering, and multimodal evaluation under each study's native protocol. The heterogeneity and tight coupling of these components make faithful reproduction challenging, limiting the applicability of existing automated frameworks designed for text-based LLM jailbreaks. To bridge this gap, we present \framework, a paper-to-pipeline agent framework for efficient and faithful reproduction of T2I jailbreak methods. \framework separately constructs paper-specific attack modules and evaluation pipelines, then integrates them through a unified runtime contract that decouples attack logic from shared evaluation infrastructure. Specification audits ensure consistency with each paper's original methodology and evaluation protocol. To further improve reproduction quality and efficiency, we introduce experience-driven self-improvement, combining execution-guided refinement within each task with persistent memory for reusing verified implementations across tasks. We evaluate \framework on 11 representative T2I jailbreak methods, achieving mean and median absolute reproduction errors of 3.3 and 1.7 percentage points, respectively, and establishing a standardized cross-model benchmark across 10 reproduced methods. Across three attack families, memory reuse improves code quality by up to 11.5\% and reduces audit iterations for two methods. Moreover, on a held-out method, progressively accumulated memory increases the reproduced attack success rate from 45.3\% to 57.0\%. These results demonstrate that \framework enables faithful automated T2I jailbreak reproduction while improving subsequent implementations through accumulated experience.
\end{abstract}

\section{Introduction}
Text-to-image (T2I) models have been widely adopted for visual content generation~\cite{rombach2022high,yang2023diffusion,Naseh_2026_CVPR}, but remain vulnerable to adversarial prompts that bypass safety safeguards~\cite{schramowski2023safe}. Understanding these vulnerabilities requires reliable reproduction and evaluation of existing T2I jailbreak methods. Unlike text-based LLM jailbreaks~\cite{yi2024jailbreak,liu2024jailbreak}, T2I attacks involve diverse mechanisms that interact with image generation, safety filtering, and multimodal evaluation~\cite{yang2024mma,yang2024sneakyprompt,jin2025jailbreakdiffbench}. Reproducing these methods therefore requires reconstructing heterogeneous attack and evaluation pipelines while preserving paper-specific configurations, including victim models, sampling strategies, safety checkers, and success criteria. Even subtle discrepancies in these components can substantially affect the reported attack success rate (ASR), making faithful reproduction challenging.

Recent work has explored automating the reproduction of jailbreak research. In particular, Jailbreak Foundry~\cite{fang2026jailbreak} translates LLM jailbreak papers into executable attack modules under a unified framework. However, its design primarily targets text-based attacks and does not directly accommodate the heterogeneous generation and evaluation pipelines required by T2I jailbreaks. Extending such frameworks beyond text-only attacks requires more than adapting attack algorithms: it demands synchronizing multi-stage workflows—spanning prompt transformation, diffusion sampling, image-level filtering, and VLM judging—under a unified execution interface. This gap motivates a dedicated paper-to-pipeline framework for automated T2I jailbreak reproduction.

To bridge this gap, we propose \textbf{\framework}, a paper-to-pipeline agent framework for efficient and faithful reproduction of T2I jailbreak methods. Given a research paper and optional reference code, \framework separately constructs paper-specific attack modules and evaluation pipelines, integrating them through a unified runtime contract. By decoupling attack logic from shared evaluation infrastructure, \framework accommodates diverse attack mechanisms while preserving each paper's original methodology and evaluation protocol.
To further improve reproduction quality and efficiency, we introduce \textbf{experience-driven self-improvement}. Within each task, specification audits and execution feedback guide iterative refinement of generated implementations. Across tasks, \textsc{CIPHER-Memory} accumulates verified implementations and reuses them to guide subsequent reproductions. Retrieved experience is adapted and re-audited against the target paper's specifications to preserve reproduction fidelity.

Using \framework, we reproduce eleven representative T2I jailbreak methods under their original experimental protocols and evaluate ten of them under a unified cross-model benchmark.
Our main contributions are summarized as follows:
\begin{itemize}
    \item We present \textbf{\framework}, a paper-to-pipeline agent framework that decouples paper-specific attack logic from shared evaluation infrastructure under a unified runtime contract, enabling efficient and faithful reproduction of diverse T2I jailbreak methods.

    \item We introduce \textbf{experience-driven self-improvement}, combining specification- and execution-guided refinement within tasks with audit-gated memory for reusing verified implementations across tasks. Memory reuse improves code quality by up to 11.5\% and yields an 11.7-percentage-point ASR improvement on a held-out method as experience accumulates.

    \item We reproduce eleven representative T2I jailbreak methods, including those without public code, achieving a mean absolute reproduction error of 3.3 percentage points. We further evaluate ten reproduced methods under a unified cross-model protocol, enabling systematic comparisons of T2I jailbreak effectiveness.
\end{itemize}

\section{Related Work}
\noindent\textbf{Jailbreak Attacks on Text-to-Image Models.}
T2I jailbreaks encompass token-level perturbations, gradient- or query-based adversarial optimization, and agentic search. SneakyPrompt employs reinforcement learning to bypass safety filters~\cite{yang2024sneakyprompt}, MMA-Diffusion crafts multimodal perturbations against text and image safety checkers~\cite{yang2024mma}, and Ring-A-Bell probes concept-erasure defenses via representation-level optimization~\cite{tsai2024ring}. More recently, JailFuzzer introduces feedback-driven prompt mutation~\cite{dong2025fuzz}, PromptTune fine-tunes an attacker LLM for query-free transfer~\cite{jiang2026jailbreaking}, and ART~\cite{li2024art} coordinates LLMs and vision-language models (VLMs) to uncover unsafe generations triggered by seemingly benign prompts. Because these methods intertwine prompt rewriting, iterative search, and multimodal feedback~\cite{villa2025exposing}, evaluating them requires pipeline-level reconstruction that captures cross-component interactions.

\begin{figure*}[t]
    \centering
    \makeatletter
    \let\Gread@transgrouptrue\Gread@transgroupfalse
    \makeatother
    \includegraphics[width=0.95\linewidth]{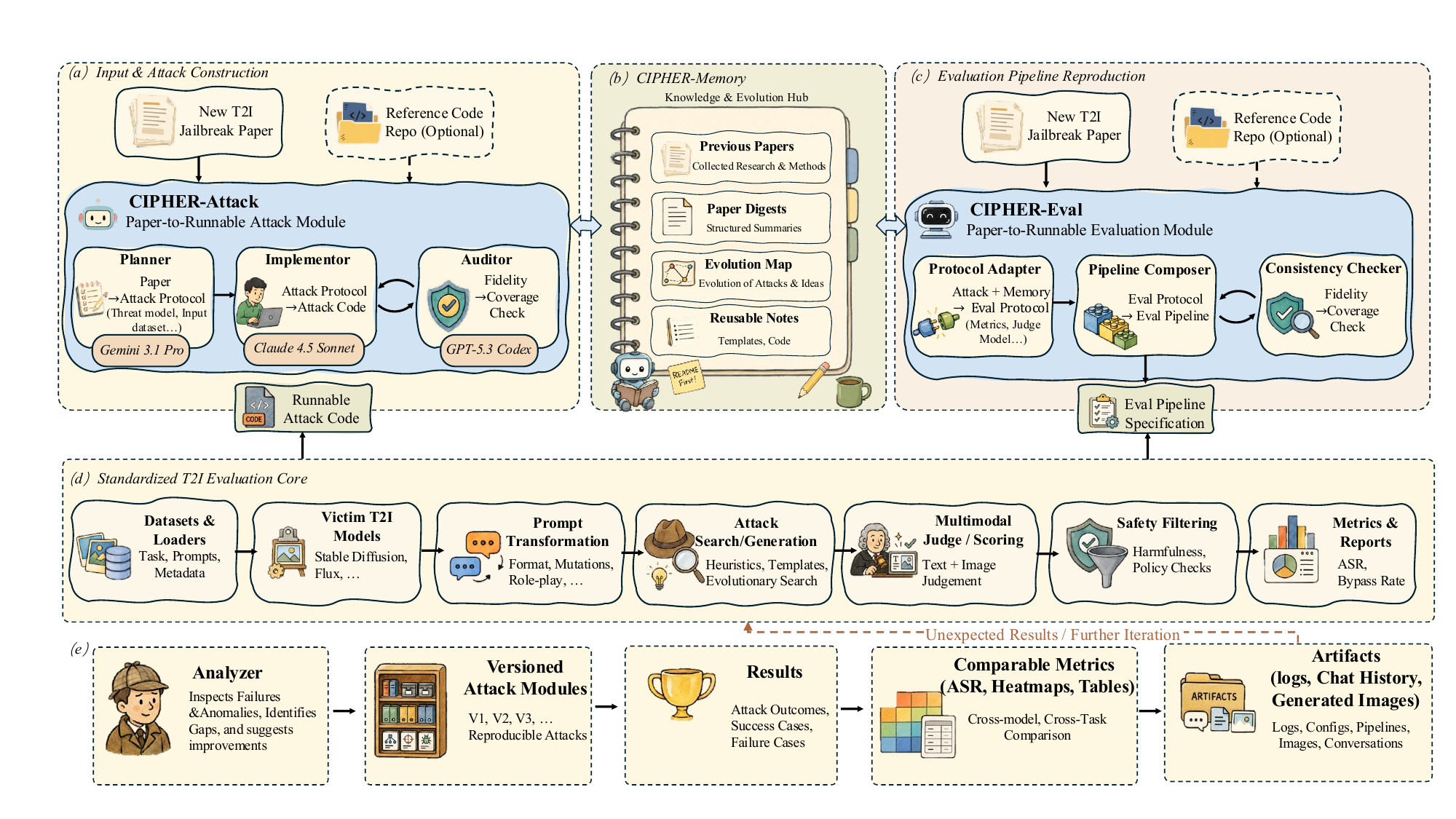}
    \caption{\textbf{The framework of \framework.} Given a research paper and optional reference code, \framework constructs an executable attack module and evaluation pipeline under a unified runtime contract. A shared evaluation core supports standardized benchmarking, while specification audits and execution feedback refine implementations. Verified implementations are stored in memory for reuse across tasks.}
    \label{fig:framework}
\end{figure*}

\noindent\textbf{Automated Paper-to-Code Reproduction.}
Recent scientific agents synthesize executable code from research manuscripts~\cite{lu2024ai,siegel2024core,schmidgall2025agent,seo2025paper2code}, while benchmarks such as PaperBench~\cite{starace2025paperbench} and MLE-bench~\cite{chan2024mlebench} evaluate replication fidelity and end-to-end machine learning engineering. Jailbreak Foundry specializes in text-based LLM attacks~\cite{fang2026jailbreak}, whereas T2I reproduction additionally demands precise alignment across image synthesis, visual-feature evasion, safety filters, and multimodal judges. \framework instead constructs attack and evaluation modules under a shared runtime contract, preserving each paper's threat model, query budget, and success criteria.

\noindent\textbf{Self-Improving Agents.}
Self-improvement mechanisms let agents refine their capabilities over time: Reflexion buffers verbal self-critique~\cite{shinn2023reflexion}, STOP optimizes code-generation scaffolds~\cite{zelikman2023stop}, the Darwin G\"odel Machine evolves coding agents via execution verification~\cite{zhang2025darwin}, and A-MEM~\cite{xu2025amem} organizes historical experience through structured notes and relational links.

\section{\framework Framework}
As illustrated in \Cref{fig:framework}, \framework automatically reproduces T2I jailbreak methods by constructing executable attack modules and evaluation pipelines under a unified runtime contract. Our framework incorporates \textbf{experience-driven self-improvement} through two complementary mechanisms: \emph{within-task refinement} and \emph{cross-task experience reuse}. Within each task, specification audits and execution feedback guide iterative refinement of generated implementations. Across tasks, verified implementations are accumulated in \textsc{CIPHER-Memory} and reused to guide subsequent reproductions, with fidelity to each target paper maintained through specification audits.

\subsection{Task Formulation}

Given a jailbreak paper $p$, an optional reference repository $R$, a unified module contract $C$, and the memory $\mathcal{M}$ (\textsc{CIPHER-Memory}), \framework synthesizes two executable artifacts: an attack module $m_p$ and an evaluation pipeline $e_p$. The attack module $m_p$ instantiates the core mechanism described in $p$, mapping harmful intents to adversarial prompts or search trajectories. The evaluation pipeline $e_p$ executes the end-to-end assessment protocol across designated victim models, safety filters, judges, and metrics.

The input manuscript is first normalized into structured Markdown, \mbox{$x_p=\mathrm{NormalizeToMD}(p)$}. From this representation, \framework generates formal attack and evaluation specifications and compiles them into executable artifacts:
\begin{align}
s_p^{\mathrm{atk}} &= \pi_{\mathrm{atk}}(x_p,C,R,\mathcal{M}), \\
m_p &= \kappa_{\mathrm{atk}}(s_p^{\mathrm{atk}},C,R), \\
s_p^{\mathrm{eval}} &= \pi_{\mathrm{eval}}(x_p,m_p,C,R,\mathcal{M}), \\
e_p &= \kappa_{\mathrm{eval}}(s_p^{\mathrm{eval}},C).
\end{align}
Here, $\pi_{\mathrm{atk}}$ and $\pi_{\mathrm{eval}}$ are planning operators that yield structured specifications, and $\kappa_{\mathrm{atk}}$ and $\kappa_{\mathrm{eval}}$ are implementation operators that instantiate executable modules conforming to $C$.

\mypara{Experience State and Reuse}
Let $\mathcal{M}_k$ denote the memory before reproducing paper $p_k$, and let $h_k=\mathrm{Retrieve}(\mathcal{M}_k,p_k)$ be the experience retrieved for construction. After synthesis and verification, a candidate record $d_k$ stores the accepted attack implementation's file path, functional summary, structural fingerprint, and reuse lineage. The cross-task memory state updates as
\begin{equation}
\mathcal{M}_{k+1}=\begin{cases}
\mathcal{M}_k\oplus d_k, & a_k^{\mathrm{atk}}=1,\\
\mathcal{M}_k, & a_k^{\mathrm{atk}}=0,
\end{cases}
\label{eq:rsi-memory}
\end{equation}
where $a_k^{\mathrm{atk}}\in\{0,1\}$ indicates whether the module passes both attack-audit verification and readiness checks, and $\oplus$ appends the structured record to memory. Subsequent reproductions retrieve context from $\mathcal{M}_{k+1}$. Here, index $t$ tracks within-task refinement iterations, whereas index $k$ tracks cross-task experience accumulation.

\mypara{Unified Contract and Runtime Core}
\framework enforces a unified runtime contract $C=(\mathcal{X},\Theta,\mathcal{Y},\mathcal{A})$ across all synthesized attack modules and evaluation pipelines, specifying the input schema $\mathcal{X}$, typed parameter space $\Theta$, output schema $\mathcal{Y}$, and auditable artifact store $\mathcal{A}$.

In T2I jailbreak tasks, $\mathcal{X}$ holds task instructions, target harmful prompts, metadata, and optional random seeds; $\Theta$ defines attack and search budgets, sampler configurations, judge endpoints, and safety filters; $\mathcal{Y}$ records perturbed prompts, synthesized images, and safety judgments; and $\mathcal{A}$ archives execution logs, configurations, intermediate states, and evaluation reports. This abstraction decouples paper-specific attack logic from shared evaluation infrastructure, enabling plug-and-play cross-method benchmarking.

\subsection{\textsc{CIPHER-Attack}: From Paper to Runnable Attack Module}

The attack construction stage compiles a jailbreak manuscript into an executable module under contract $C$ via a closed Planner--Implementor--Auditor loop.

The \textbf{Planner} analyzes the normalized text $x_p$, optional reference code $R$, and retrieved memory records to produce the specification $s_p^{\mathrm{atk}}$. It extracts the optimization objective, threat model, perturbation operator, search dynamics, stopping criteria, query budget, and hyperparameters, and binds the method to the standardized \texttt{generate\_attack()} entry point.

The \textbf{Implementor} translates $s_p^{\mathrm{atk}}$ into the executable module $m_p$, instantiating method-specific logic under contract $C$. Algorithmic hyperparameters are exposed through the typed parameter space $\Theta$, whereas data ingestion, victim-model invocation, artifact caching, and scoring are delegated to the shared runtime core.

Finally, the \textbf{Auditor} verifies alignment among $m_p$, $s_p^{\mathrm{atk}}$, $C$, and any reference code $R$:
\begin{equation}
(a_t^{\mathrm{atk}},r_t^{\mathrm{atk}})=\alpha_{\mathrm{atk}}(m_p,s_p^{\mathrm{atk}},C,R),
\end{equation}
where $a_t^{\mathrm{atk}}\in\{0,1\}$ denotes audit acceptance at iteration $t$, and $r_t^{\mathrm{atk}}$ is an actionable diagnostic report if discrepancies arise. The Auditor inspects control flow, prompt templates, default hyperparameters, query budgets, search termination rules, and core algorithmic operators against the source materials. Whenever $a_t^{\mathrm{atk}}=0$, the Implementor refines $m_p$ guided by $r_t^{\mathrm{atk}}$ until the module passes verification or reaches the iteration budget $T$.

\subsection{\textsc{CIPHER-Eval}: From Paper to Runnable Evaluation Pipeline}

Because T2I jailbreak papers employ diverse victim models, prompt-evolution strategies, safety filters, and judging criteria, \framework builds the evaluation pipeline in a dedicated stage whose roles mirror those above: the \textbf{Protocol Adapter} plays the Planner, the \textbf{Pipeline Composer} the Implementor, and the \textbf{Consistency Checker} the Auditor. The Protocol Adapter parses the manuscript to produce $s_p^{\mathrm{eval}}$, specifying target datasets, victim architectures, sampler hyperparameters, safety checkers, judges, and metrics. The Pipeline Composer then wires the verified attack module $m_p$ to standardized data loaders, model adapters, judge interfaces, and logging utilities.

The Consistency Checker audits whether the assembled pipeline $e_p$ faithfully instantiates the evaluation specification:
\begin{equation}
(a_t^{\mathrm{eval}},r_t^{\mathrm{eval}})=\alpha_{\mathrm{eval}}(e_p,s_p^{\mathrm{eval}},C),
\end{equation}
where $a_t^{\mathrm{eval}}\in\{0,1\}$ indicates verification success at iteration $t$, and $r_t^{\mathrm{eval}}$ contains targeted revision feedback. This check validates victim-model configurations, filter hooks, judge prompts, and metric aggregation against both $s_p^{\mathrm{eval}}$ and the source paper. If verification fails, the Pipeline Composer revises $e_p$ using $r_t^{\mathrm{eval}}$ within budget $T$.

Decoupling attack synthesis from evaluation construction lets \framework support both \emph{paper-matched reproduction} (validating fidelity under each paper's native setup) and \emph{standardized evaluation} (comparing relative effectiveness under a unified benchmark).

Algorithm~\ref{alg:cipher_short} summarizes the end-to-end workflow. When the empirical discrepancy $|\Delta_p|$ exceeds tolerance $\tau$, the Analyzer inspects execution logs and discrepancy reports to trigger \texttt{RefineAndRebuild}. This post-execution refinement is restricted to resolving engineering defects: it may not alter the core attack algorithm or tune hyperparameters to match reported numbers. It only diagnoses and repairs systems-level bugs, such as API timeouts that silently drop samples, image-decoding format mismatches, or version discrepancies in safety-filter threshold mappings.

\begin{algorithm}[t!]
\caption{\framework: Artifact Refinement and Experience Reuse}
\label{alg:cipher_short}
\begin{algorithmic}[1]
\REQUIRE Paper $p_k$, contract $C$, memory $\mathcal{M}_k$, audit budget $T$, discrepancy tolerance $\tau$

\STATE $p \leftarrow p_k$, $\mathcal{M}_{k+1}\leftarrow\mathcal{M}_k$
\STATE $x_p \leftarrow \mathrm{NormalizeToMD}(p)$, $R \leftarrow \mathrm{RetrieveRepo}(p)$, $h_p \leftarrow \mathrm{RetrieveMemory}(\mathcal{M}_k,p)$
\STATE $m_p \leftarrow \mathrm{ForgeAttack}(x_p,C,R,h_p,T)$
\STATE $e_p \leftarrow \mathrm{ForgeEval}(x_p,m_p,C,R,h_p,T)$

\STATE $\mathcal{A}_p \leftarrow \mathrm{RunMatchedEval}(m_p, e_p)$
\STATE $\Delta_p \leftarrow \mathrm{ASR}_{\mathrm{paper}}-\mathrm{ComputeASR}(\mathcal{A}_p)$

\IF{$|\Delta_p| > \tau$}
    \STATE $\begin{aligned}[t]
        &(m_p', e_p') \leftarrow \mathrm{RefineAndRebuild}(\\
        &\qquad x_p,m_p,e_p,C,R,h_p,\mathcal{A}_p,\Delta_p)
    \end{aligned}$
    \STATE $\mathcal{A}_p' \leftarrow \mathrm{RunMatchedEval}(m_p', e_p')$
    \STATE $\Delta_p' \leftarrow \mathrm{ASR}_{\mathrm{paper}}-\mathrm{ComputeASR}(\mathcal{A}_p')$
    \IF{$|\Delta_p'| \le |\Delta_p|$}
        \STATE $m_p \leftarrow m_p'$, $e_p \leftarrow e_p'$
        \STATE $\mathcal{A}_p \leftarrow \mathcal{A}_p'$, $\Delta_p \leftarrow \Delta_p'$
    \ENDIF
\ENDIF

\IF{$\mathrm{AttackAuditAndReadiness}(m_p)=1$}
    \STATE $d_k\leftarrow\mathrm{ImplementationRecord}(p,m_p)$
    \STATE $\mathcal{M}_{k+1}\leftarrow\mathcal{M}_k\oplus d_k$
\ENDIF
\RETURN $m_p,e_p,\Delta_p,\mathcal{A}_p,\mathcal{M}_{k+1}$
\end{algorithmic}
\end{algorithm}

\mypara{Standardized Text-to-Image Evaluation Core}
Given a benchmark dataset $\mathcal{D}=\{q_i\}_{i=1}^{N}$, the attack module $m_p$ and victim generator $G_v$ sequentially produce an adversarial prompt and an image:
\begin{equation}
\tilde{q}_i=m_p(q_i;\theta_p),\qquad I_i=G_v(\tilde{q}_i),
\end{equation}
where $\theta_p\in\Theta$ is the paper-matched attack configuration. Each evaluation protocol specifies a semantic judge $J$ and a safety filter $F$. The attack success indicator for sample $i$ is
\begin{equation}
s_i=\mathbf{1}\!\left[J(q_i,\tilde{q}_i,I_i)\ge \tau_J \ \land\ F(I_i)=0\right],
\end{equation}
where $\tau_J$ is the judge's decision threshold and $F(I_i)=0$ indicates that the image evades the target safety filter. Overall ASR averages $s_i$ across valid samples (multi-attempt extensions are given in the Metrics paragraph), and our standardized benchmark adopts the image-only GPT-4o judging criterion in~\Cref{sec:evaluation-setup}.

Under the paper-matched setting, let $\mathrm{ASR}_{\mathrm{paper}}$ denote the metric reported in the original publication. We quantify reproduction discrepancy as
\begin{equation}
\Delta_p=\mathrm{ASR}_{\mathrm{paper}}-\mathrm{ASR}_{\mathrm{gen}},
\end{equation}
where $\mathrm{ASR}_{\mathrm{gen}}$ is obtained by executing our synthesized pipeline under the paper-matched protocol. The standardized setting instead fixes the benchmark dataset and assessment protocol across all methods, yielding a cross-method, cross-victim ASR matrix $\mathbf{A}_{p,v}$ for direct comparison.

\subsection{Audit-Gated Experience Reuse}

\textsc{CIPHER-Memory} links successive reproductions through the audit-gated state update in~\Cref{eq:rsi-memory}. Each committed record stores algorithmic metadata, file paths, structural fingerprints, audit coverage metrics, and reuse lineage, providing summaries and code templates that guide subsequent planning, adaptation, and composition.

\begin{figure*}[ht]
    \centering
    \includegraphics[width=0.95\linewidth]{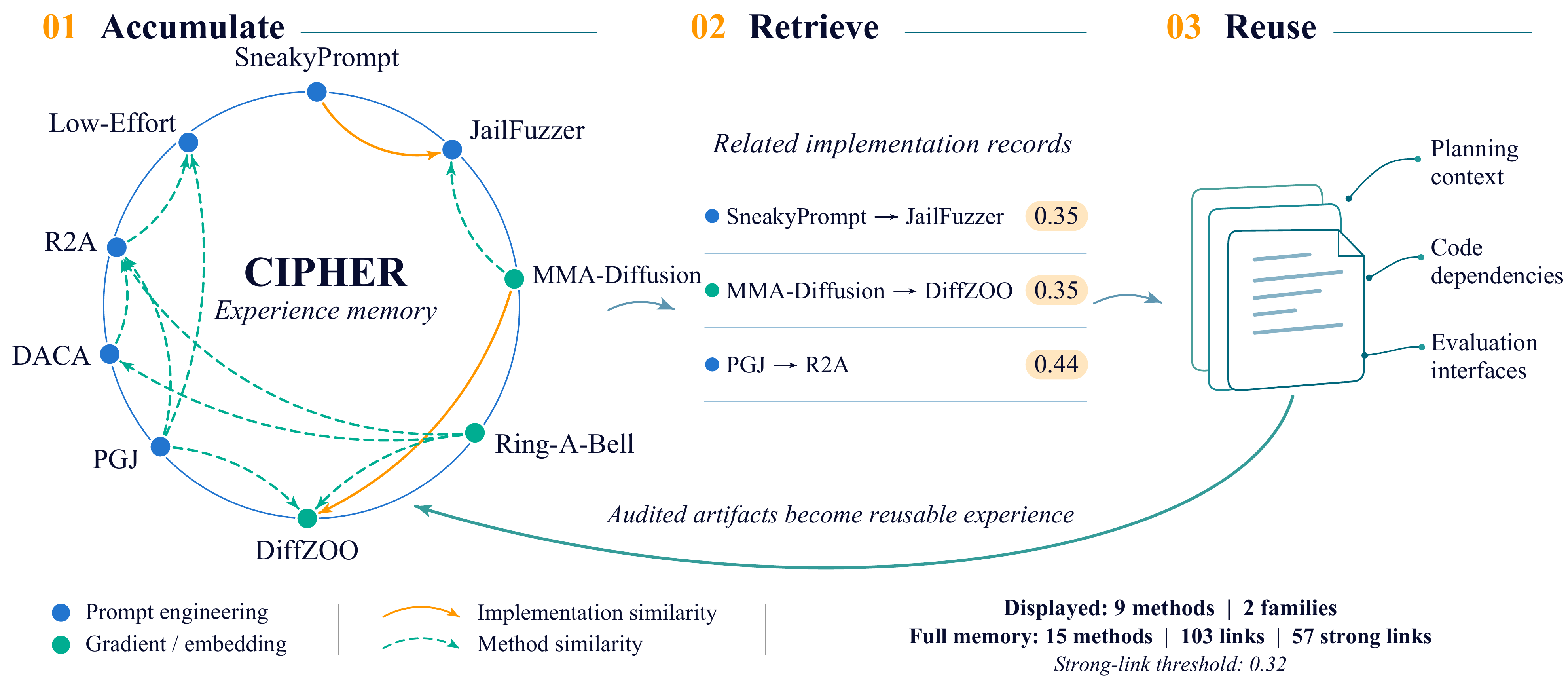}
    \caption{\textbf{Relational graph of audited reproduction records in \textsc{CIPHER-Memory}}, illustrating nine reproduced methods across two primary paradigms along with typed edges that encode implementation reuse and methodological similarity.}
    \label{fig:Memory}
\end{figure*}

When constructing a new pipeline, the Planner aligns retrieved records with the target paper's methodology, the Implementor adapts reusable scaffolding to method-specific requirements, and the Auditor verifies compliance with both the source manuscript and contract $C$. Only implementations passing all verification gates enter memory with full provenance. Candidate records are ranked via token overlap and agent-variant metadata, while the relational graph in~\Cref{fig:Memory} captures structural and methodological affinities across attack families (see~\Cref{app:memory_scoring} for retrieval and graph-scoring formulations).

\subsection{Comparison with Existing Reproduction Frameworks}

Existing frameworks support general paper-to-code generation or automated reproduction of text-based jailbreaks. As shown in Table~\ref{tab:framework_comparison}, \framework extends automated reproduction to multimodal T2I jailbreaks by reconstructing both attack modules and paper-specific evaluation pipelines. Furthermore, it enables cross-task experience-driven improvement by accumulating and reusing verified implementations.

\begin{table}[t]
    \centering
    \setlength{\tabcolsep}{1pt}
    \renewcommand{\arraystretch}{1}
    \small
    \begin{tabular}{lccc}
        \toprule
        \textbf{Capability}
        & \textbf{Paper2Code \cite{seo2025paper2code}}
        & \textbf{JBF \cite{fang2026jailbreak}}
        & \textbf{Ours} \\
        \midrule
        Paper-to-code generation
        & $\checkmark$ & $\checkmark$ & $\checkmark$ \\
        Jailbreak reproduction
        & -- & $\checkmark$ & $\checkmark$ \\
        Multimodal execution
        & -- & -- & $\checkmark$ \\
        Paper-specific eval. synthesis
        & -- & -- & $\checkmark$ \\
        Standardized benchmarking
        & -- & $\checkmark$ & $\checkmark$ \\
        \shortstack[l]{Experience-driven\\improvement}
        & -- & -- & $\checkmark$ \\
        \bottomrule
    \end{tabular}
    \caption{\textbf{Comparison with existing reproduction frameworks.}
    Checkmarks indicate explicitly supported capabilities, while dashes denote capabilities not explicitly provided.}
    \label{tab:framework_comparison}
\end{table}

\section{Experiments}

\mypara{Method Coverage and Selection}
We evaluate \framework on eleven representative T2I jailbreak methods spanning token-level substitution, semantic rewriting, gradient/query-based adversarial optimization, and LLM/VLM-driven agentic search. These methods cover white-box, black-box, query-intensive, and transfer-based threat models targeting text filters, image-level safety checkers, and concept-erasure defenses. Supplementary~\Cref{tab:attack-taxonomy} details their transformation operators and search strategies under our unified contract.

\begin{table*}[t]
\centering
\small
\resizebox{\textwidth}{!}{\begin{tabular}{@{}lcccccccc@{}}
\toprule
\multirow{2}{*}{\textbf{Paper (YYMM) / Attack}}
& \multicolumn{2}{c}{\textbf{Evaluation Setup}}
& \multicolumn{3}{c}{\textbf{Implementation}}
& \multicolumn{3}{c}{\textbf{Eval Metrics}} \\
\cmidrule(lr){2-3} \cmidrule(lr){4-6} \cmidrule(l){7-9}
& Dataset
& Metric
& Code Ref
& Victim Model
& $\rho$
& $\mathrm{ASR}_{\mathrm{paper}}$ (\%)
& $\mathrm{ASR}_{\mathrm{gen}}$ (\%)
& $\Delta$ (pp) \\
\midrule
23-05 / SneakyPrompt~\cite{yang2024sneakyprompt} & NSFW-200 & Bypass Rate & YES & SDv1.4 & $-1075$ & 100 & 100 & 0.0 \\
23-11 / MMA-Diffusion~\cite{yang2024mma} & MMA & ASR-1 & YES & SDv1.5 & $-619$ & 54.2 & 50.0 & 4.2 \\
23-12 / DACA~\cite{deng2024harnessingllmattackllmguardedDACA} & VBCDE & ASR-1 & YES & SDv1.4 & $+42$ & 30.0 & 28.3 & 1.7 \\
24-03 / Ring-A-Bell~\cite{tsai2024ring} & I2P & ASR-1 & YES & SLD & $+72$ & 93.7 & 90.5 & 3.2 \\
24-04 / JPA~\cite{ma2024jailbreaking} & I2P & ASR-1 & YES & SLD & $+33$ & 90.9 & 91.1 & $-0.2$ \\
24-08 / JailFuzzer~\cite{dong2025fuzz} & NSFW-200 & Bypass Rate & YES & SDv1.4 & $-1309$ & 100 & 100 & 0.0 \\
24-12 / HTS~\cite{gao2024htsattackheuristictokensearch} & LAION-COCO & Bypass Rate & NO & SDv1.4 & -- & 81.5 & 89.4 & $-7.9$ \\
25-02 / DiffZOO~\cite{dang2025diffzoopurelyquerybasedblackbox} & I2P & ASR-1 & YES & SDv1.4 & $-877$ & 59.0 & 59.6 & $-0.6$ \\
25-02 / PGJ~\cite{huang2025perceptionguidedjailbreaktexttoimagemodels} & NSFW-200 & ASR-1 & NO & SDXL & -- & 100 & 92.8 & 7.2 \\
25-03 / R2A~\cite{zhang2025reason2attackjailbreakingtexttoimagemodels} & I2P & ASR-1 & NO & SDv1.4 & -- & 90.0 & 79.1 & 10.9 \\
26-04 / Low-Effort~\cite{mustafa2026loweffortjailbreakattackstexttoimage} & I2P & ASR-4 & NO & SDv1.5 & -- & 71.5 & 71.8 & $-0.3$ \\
\bottomrule
\end{tabular}}
\caption{\textbf{Overview and paper-matched reproduction results across eleven representative T2I jailbreak methods.} $\rho$ denotes the difference in lines of code (LoC) between our synthesized module and the original repository. We report paper-reported ($\mathrm{ASR}_{\mathrm{paper}}$) and reproduced ($\mathrm{ASR}_{\mathrm{gen}}$) metrics, along with the signed discrepancy $\Delta=\mathrm{ASR}_{\mathrm{paper}}-\mathrm{ASR}_{\mathrm{gen}}$ in percentage points (pp).}
\label{tab:jailbreak-overview}
\end{table*}

\subsection{Data, Models, and Metrics}\label{sec:evaluation-setup}
\mypara{Data and Models}
Paper-matched evaluations follow each study's native victim configurations, including SDv1.4/v1.5~\cite{rombach2022high}, SLD~\cite{schramowski2023safe}, and SDXL~\cite{podell2023sdxlimprovinglatentdiffusion}, evaluated on NSFW-200~\cite{yang2024sneakyprompt}, MMA~\cite{yang2024mma}, VBCDE~\cite{deng2024harnessingllmattackllmguardedDACA}, I2P~\cite{schramowski2023safe}, and LAION-COCO (\Cref{tab:jailbreak-overview}). For the standardized benchmark in~\Cref{tab:standardized_benchmark}, each valid NSFW-200 prompt is allocated one attempt per model ($K=1$). The same image-only GPT-4o judge evaluates all generated images. The denominator $N_{\mathrm{valid}}$ excludes only inputs identified as invalid before execution. Prompt refusals, policy blocks, timeouts, API errors, and generation failures count as unsuccessful attempts and remain in the denominator.

\mypara{Metrics}
Given $K$ attack attempts per target prompt $q_i$, let $s_{i,a}\in\{0,1\}$ denote whether the $a$-th attempt succeeds. We compute $\mathrm{ASR}\text{-}K$ over $N_{\mathrm{valid}}$ valid prompts as
\begin{equation}
\mathrm{ASR}\text{-}K =
\frac{1}{N_{\mathrm{valid}}}
\sum_{i=1}^{N_{\mathrm{valid}}}
\mathbf{1}\left[\sum_{a=1}^{K}s_{i,a}\ge 1\right],
\end{equation}
where ASR-1 and ASR-4 correspond to $K=1$ and $K=4$. For safety-filter evasion studies, let $b_i\in\{0,1\}$ indicate whether the image synthesized for prompt $q_i$ bypasses the target filter; the bypass rate is
\begin{equation}
\mathrm{BypassRate} =
\frac{1}{N_{\mathrm{valid}}}
\sum_{i=1}^{N_{\mathrm{valid}}}
b_i .
\end{equation}
For paper-matched protocols that also enforce semantic preservation via CLIP similarity, an attempt succeeds if and only if it evades the safety filter and exceeds the similarity threshold $\delta$:
\begin{equation}
s_{i,a} =
\mathbf{1}\left[
b_{i,a}=1 \land \mathrm{CLIP}(q_i,I_{i,a})\ge \delta
\right],
\end{equation}
where $\delta=0.26$, following the original evaluation protocols.

\mypara{Implementation Details}
\label{sec:implementation_details}
We reproduce eleven T2I jailbreak methods, seven with public reference repositories and four reconstructed solely from manuscript text, preserving each study's native datasets, victim models, and safety filters. In all experiments, the role-specific backbones are fixed: Gemini-3.1-Pro as Planner, Claude-4.5-Sonnet as Implementor, GPT-5.3-Codex as Auditor, and GPT-5.5 as Analyzer, although our provider-agnostic interface readily supports alternative LLM backbones. Over the eleven reproductions, \framework achieves a 95.1\% autonomous verification pass rate with zero manual code rewrites.

\subsection{Reproduction Quality and Experience Reuse}

\mypara{Paper-Matched Reproduction Fidelity}
We evaluate whether \framework replicates published results under each paper's native setup (\Cref{tab:jailbreak-overview}). Overall, \framework achieves high reproduction fidelity across all eleven methods, with a mean absolute error (MAE) of 3.3 percentage points and a median absolute error of 1.7. SneakyPrompt and JailFuzzer exactly match their reported 100\% bypass rates, while JPA, Low-Effort, and DiffZOO deviate by only 0.2, 0.3, and 0.6 percentage points. The seven methods with public code achieve an MAE of 1.4 percentage points (maximum error 4.2), whereas the four methods without public code average 6.6.

HTS, PGJ, R2A, and Low-Effort must instead be synthesized entirely from manuscript descriptions, exposing them to undocumented engineering choices, omitted hyperparameters, and subtle judge variations. Low-Effort is reproduced within 0.3 percentage points, whereas HTS, PGJ, and R2A deviate by 7.9, 7.2, and 10.9 percentage points. These discrepancies reflect the sensitivity of T2I jailbreak pipelines to under-specified system variables, such as prompt normalization, negative prompts, random seeds, sampler schedules, safety-checker weights, and judge prompts, rather than flaws in algorithmic reconstruction. Specifically, R2A's 10.9-point gap stems from missing proprietary assets: the authors' fine-tuned attack checkpoint, chain-of-thought training corpus, and post-training weights are publicly unavailable, restricting our reproduction to the paper's zero-shot frame-semantics pathway. Similarly, PGJ's 7.2-point deviation reflects unreleased LLM configurations, internal pre-checkers, its custom GPT-4-generated 1,000-prompt dataset, and exact scoring thresholds, all of which govern substitution candidates and bypass decisions. By archiving structured logs, configurations, and intermediate outputs, \framework renders every reproduced pipeline fully transparent and helps isolate the exact sources of empirical variance.

\mypara{Memory Reuse: Paired Evaluation}
To isolate the impact of \textsc{CIPHER-Memory}, we conduct paired with-memory versus without-memory ablations on three attack paradigms: JailFuzzer (agentic fuzzing), Ring-A-Bell (genetic optimization over CLIP embeddings), and DACA (LLM-driven semantic decomposition). Both conditions share identical source manuscripts, role backbones, system prompts, runtime environments, and evaluation rubrics; the control condition simply disables retrieval of prior audit-approved records. We employ Grok-4.7 as an independent judge to score Functional Fidelity (40\%), Technical Correctness (30\%), and Reproducibility (30\%) against the rubric anchors in~\Cref{app:prompt_design}. At the systems level, memory lets new reproductions inherit verified Diffusers sampler wrappers, NSFW safety-filter hooks, and VLM-judge invocation templates. Within related attack families, memory also supplies higher-level algorithmic primitives, such as iterative prompt-mutation operators or discrete token-search loops, while control logic is adapted to each paper's specification.

As shown in~\Cref{fig:memory-reuse}, enabling \textsc{CIPHER-Memory} consistently elevates code quality: composite scores rise from 8.16 to 9.10 for JailFuzzer (+11.5\% relative), from 8.49 to 8.90 for Ring-A-Bell (+4.8\%), and from 8.40 to 8.84 for DACA (+5.2\%). Memory reuse also reduces the required audit iterations by one round for both JailFuzzer and Ring-A-Bell while maintaining parity for DACA, confirming that audited implementation experience accelerates convergence toward higher-quality artifacts.

\begin{figure}[t]
    \centering
    \begingroup
\definecolor{rsiBlue}{HTML}{0077BB}
\definecolor{rsiTeal}{HTML}{009988}
\definecolor{rsiInk}{HTML}{233448}
\definecolor{rsiGrid}{HTML}{E2E8EF}
\begin{tikzpicture}
\begin{axis}[
    width=\linewidth, height=5.00cm,
    ybar=5pt, bar width=16pt,
    xmin=-0.5, xmax=2.5, ymin=0, ymax=12.2,
    xtick={0,1,2},
    xticklabels={JailFuzzer,Ring-A-Bell,DACA},
    ytick={0,2,4,6,8,10},
    ylabel={Quality score (1--10)},
    title={\textbf{Paired memory reuse}},
    title style={font=\small,color=rsiInk,yshift=15pt},
    label style={font=\footnotesize,color=rsiInk},
    tick label style={font=\footnotesize,color=rsiInk},
    axis lines=left, axis line style={draw=rsiInk!60},
    tick align=outside, tick style={draw=rsiInk!50},
    ymajorgrids=true, grid style={draw=rsiGrid},
    legend style={at={(0.5,1.03)},anchor=south,legend columns=2,
        draw=none,fill=none,font=\footnotesize,text=rsiInk},
    legend image code/.code={\draw[#1] (0cm,-0.08cm) rectangle (0.22cm,0.08cm);},
    point meta=explicit symbolic,nodes near coords,
    nodes near coords style={font=\footnotesize,text=rsiInk,yshift=2pt},
    clip=false,
]
\addplot[draw=rsiBlue,fill=rsiBlue!12,line width=0.8pt]
    coordinates {(0,8.16)[8.16] (1,8.49)[8.49] (2,8.40)[8.40]};
\addplot[draw=rsiTeal,fill=rsiTeal!75,line width=0.8pt]
    coordinates {(0,9.10)[9.10] (1,8.90)[8.90] (2,8.84)[8.84]};
\legend{w/o memory,w/ memory}
\node[text=rsiTeal,font=\footnotesize\bfseries] at (axis cs:0,11.1) {11.5\% gain};
\node[text=rsiTeal,font=\footnotesize\bfseries] at (axis cs:1,11.1) {4.8\% gain};
\node[text=rsiTeal,font=\footnotesize\bfseries] at (axis cs:2,11.1) {5.2\% gain};
\node[text=rsiInk,font=\footnotesize,anchor=north,yshift=-17pt] at (axis cs:0,0) {Audits: $3\to2$};
\node[text=rsiInk,font=\footnotesize,anchor=north,yshift=-17pt] at (axis cs:1,0) {Audits: $3\to2$};
\node[text=rsiInk,font=\footnotesize,anchor=north,yshift=-17pt] at (axis cs:2,0) {Audits: $2\to2$};
\end{axis}
\end{tikzpicture}
\endgroup
    \caption{\textbf{Paired ablation of \textsc{CIPHER-Memory} reuse across three attack families.} Bars report weighted code-quality scores alongside relative percentage gains; bottom annotations indicate audit rounds to acceptance (without memory $\to$ with memory).}
    \label{fig:memory-reuse}
\end{figure}

\mypara{Memory Reuse: Longitudinal Evaluation}
We evaluate cumulative reuse by reproducing a held-out method, ICER~\cite{chin2026icer}, across four memory checkpoints: $\mathcal{M}_0$, $\mathcal{M}_3$, $\mathcal{M}_6$, and $\mathcal{M}_9$. Starting from $\mathcal{M}_0=\emptyset$, we build later checkpoints by committing audited records from nine other papers, ordered by publication chronology and methodological lineage. Role models, prompts, and workflow budgets are fixed across the four runs. Each run synthesizes a fresh ICER implementation conditioned only on its checkpoint (with all ICER-specific records excluded to prevent leakage) and evaluates it against SLD-MAX under ICER's paper-matched I2P protocol.

\begin{figure}[t]
    \centering
    \includegraphics[width=0.4\textwidth]{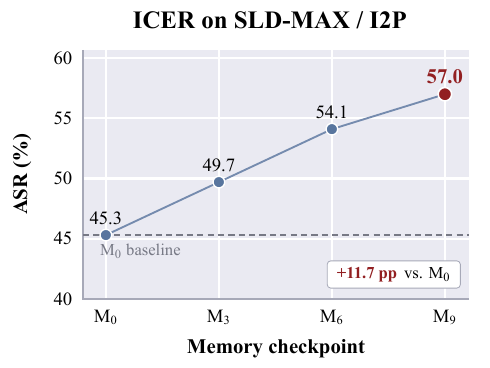}
    \caption{\textbf{Longitudinal evaluation of cumulative memory reuse on held-out ICER.} Each run synthesizes a fresh ICER pipeline conditioned on checkpoints $\mathcal{M}_0 \to \mathcal{M}_9$ accrued from other papers (excluding any ICER records), evaluated on SLD-MAX under the paper-matched I2P setup.}
    \label{fig:icer-rsi}
\end{figure}

As shown in~\Cref{fig:icer-rsi}, downstream ASR increases monotonically as memory accumulates, from 45.3\% at $\mathcal{M}_0$ to 49.7\% ($\mathcal{M}_3$), 54.1\% ($\mathcal{M}_6$), and 57.0\% ($\mathcal{M}_9$). The final checkpoint yields an 11.7-percentage-point absolute improvement (+25.8\% relative) over the memoryless baseline, demonstrating that cross-task engineering experience from earlier reproductions substantially improves execution performance on unseen target papers.

\begin{figure*}[t]
    \centering
    \includegraphics[width=\textwidth]{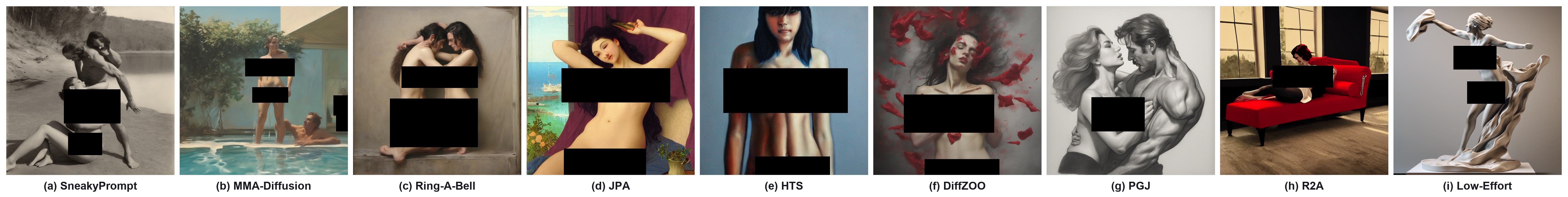}
    \caption{\textbf{Qualitative examples synthesized by nine reproduced attack pipelines under their respective evaluation settings.} Sensitive visual regions are redacted with black bars.}
    \label{fig:qualitative-attacks}
\end{figure*}

\begin{figure}[!ht]
    \centering
    \includegraphics[width=\linewidth]{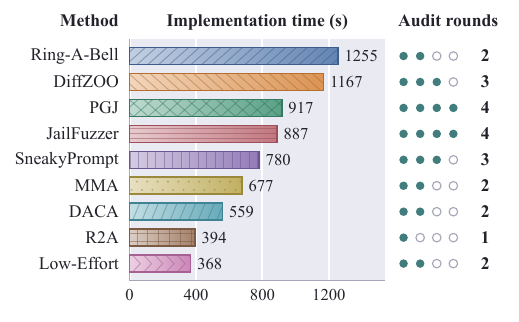}
    \caption{\textbf{Synthesis wall-clock time and audit iterations across the nine-method efficiency subset.} Blue bars denote total duration in seconds; filled teal dots indicate the number of audit rounds required for acceptance.}
    \label{fig:implementation_time}
\end{figure}

\mypara{External and Single-Pass Baselines}
We also benchmark \framework against external paper-to-code and single-pass coding baselines under matched conditions: Paper2Code on Ring-A-Bell, and single-pass Cursor-agent runs (powered by Claude-4.5-Sonnet) on DACA and R2A. We assess module-level alignment with the source paper and require a runnable, paper-aligned pipeline for end-to-end validation. Two of the three baseline runs passed the module-level check, and zero passed end-to-end validation. \framework passed both checks in all three matched comparisons.

\mypara{Standardized Benchmark across T2I Victim Models}
Prior T2I jailbreak studies evaluate on disparate datasets, victim models, and judges, so their reported metrics cannot be directly compared. We therefore benchmark the ten reproduced methods whose paper-matched discrepancy satisfies $|\Delta|\le 10$ percentage points (all but R2A; \Cref{tab:jailbreak-overview}) under a unified protocol. By binding their \framework-synthesized attack modules to our shared evaluation core, we fix the dataset (NSFW-200), metric (ASR-1), safety filters, and multimodal judge (image-only GPT-4o; see~\Cref{sec:evaluation-setup}) across four diverse victim models: SDv1.4~\cite{rombach2022high}, SDv1.5, SDXL~\cite{podell2023sdxlimprovinglatentdiffusion}, and GPT-image-2~\cite{openai2026images2}.

As shown in~\Cref{tab:standardized_benchmark}, attack effectiveness varies sharply across method families and victim architectures. On open-weights diffusion models, DACA outperforms all other methods, achieving 94.5\%, 95.0\%, and 96.7\% ASR on SDv1.4, SDv1.5, and SDXL. In contrast, SneakyPrompt, Ring-A-Bell, and JailFuzzer range from 49.7\% to 67.4\%, revealing that high filter-bypass rates under paper-specific setups do not necessarily translate into high ASR under multimodal judging. This divergence highlights an insight for T2I safety evaluation: token-level substitutions and heuristic perturbations frequently evade shallow keyword or CLIP-based filters at the cost of corrupting prompt semantics, yielding distorted or ambiguous images that fail VLM harmfulness verification. Semantic decomposition attacks like DACA instead preserve the visual intent while circumventing text-level guardrails, posing a substantially higher real-world risk on open diffusion backbones. Meanwhile, the commercial closed-source GPT-image-2 shows strong robustness: all ten attacks fall below 3.5\% ASR, with SneakyPrompt, DiffZOO, and PGJ reduced to 0.0\%. Qualitative generations for nine reproduced methods are shown in~\Cref{fig:qualitative-attacks}.

\mypara{Reproduction Efficiency}
As shown in~\Cref{fig:implementation_time}, \framework produces verified pipelines in an average of 2.56 audit iterations and 778 seconds across nine methods. R2A and Low-Effort finish within 6--7 minutes, whereas search-intensive Ring-A-Bell and DiffZOO require longer synthesis and verification of custom optimization loops.

\begin{table}[t]
\centering
\small
\setlength{\tabcolsep}{3pt}
\renewcommand{\arraystretch}{1.08}
\begin{tabularx}{\linewidth}{@{}Xrrrr@{}}
\toprule
\multirow{2}{*}{\textbf{Method}} & \multicolumn{4}{c}{\textbf{ASR (\%) across victim models}} \\
\cmidrule(r){2-5}
 & \textbf{SDv1.4} & \textbf{SDv1.5} & \textbf{SDXL} & \textbf{GPT-image-2} \\
\midrule
SneakyPrompt & 53.0 & 54.7 & 58.6 & 0.0 \\
MMA-Diffusion & 60.3 & 59.7 & 52.7 & 0.8 \\
DACA & \textbf{94.5} & \textbf{95.0} & \textbf{96.7} & 2.4 \\
Ring-A-Bell & 49.7 & 50.3 & 60.2 & 3.1 \\
JailFuzzer & 50.3 & 55.2 & 67.4 & 1.1 \\
DiffZOO & 63.0 & 59.1 & 66.9 & 0.0 \\
PGJ & 63.5 & 69.6 & 86.7 & 0.0 \\
Low-Effort & 62.2 & 63.1 & 91.1 & 1.9 \\
JPA & 49.7 & 50.0 & 53.0 & \textbf{3.3} \\
HTS & 64.1 & 64.1 & 75.7 & 2.7 \\
\bottomrule
\end{tabularx}
\caption{\textbf{Standardized cross-model ASR-1 (\%) on NSFW-200 for the ten reproduced methods with paper-matched discrepancy $|\Delta|\le 10$ percentage points.} Bold values denote the highest ASR per victim model.}
\label{tab:standardized_benchmark}
\end{table}

\section{Conclusion}
We present \framework, an experience-driven paper-to-pipeline framework that couples within-task audit refinement with cross-task reuse of verified implementation records. \framework faithfully reproduces eleven representative T2I jailbreak methods under their paper-matched setups and establishes a standardized ten-method benchmark across four victim models. Paired ablations show that \textsc{CIPHER-Memory} improves code quality by 4.8\%--11.5\% while reducing audit overhead, and a longitudinal study on held-out ICER shows downstream ASR rising from 45.3\% to 57.0\% across successive memory checkpoints. Together, these findings demonstrate that structured accumulation of implementation experience significantly advances automated scientific reproduction and standardized T2I safety auditing.

\section{Ethical Considerations}
\label{sec:ethical_considerations}
\framework is designed to advance reproducible T2I safety auditing and red-teaming within controlled research environments. By standardizing evaluation across diverse attack paradigms, we aim to help model developers and safety researchers diagnose vulnerabilities and harden multimodal safeguards. All sensitive visual regions in our qualitative figures are redacted, and our responsible artifact release protocol is detailed in~\Cref{sec:appendix_artifacts}.

\bibliographystyle{plain}
\bibliography{custom}

\begin{appendices}
\section{Prompt Design}
\label{app:prompt_design}
To ensure an unbiased, rigorous, and reproducible evaluation of the code generated from research papers, we developed a structured evaluation protocol for the Large Language Model (LLM). Rather than allowing the LLM to assign arbitrary scores, we implement an \textbf{Anchor-based Scoring Rubric} across three fundamental technical dimensions.

This multi-dimensional scoring mechanism ensures that functional alignment with mathematical formulas, programmatic correctness, and experimental reproducibility are decoupled and evaluated objectively.

\subsection{Evaluation Criteria and Scoring Anchors}
Table~\ref{tab:scoring_rubric} outlines the specific evaluation criteria and the calibrated behavioral anchors used by the LLM to determine scores from 1 to 10.

\begin{table*}[htbp]
\centering
\small
\caption{Multi-Dimensional Alignment Rubric and Score Anchors}
\label{tab:scoring_rubric}
\begin{tabularx}{\textwidth}{>{\raggedright\arraybackslash}p{3cm} c X}
\toprule
\textbf{Dimension} & \textbf{Score Range} & \textbf{Behavioral Anchor Description} \\ \midrule
 & 9--10 (Perfect) & Every core attack operator, optimization objective, and search stopping criterion matches the paper specification exactly. \\
\textbf{Functional Fidelity} & 7--8 (Good) & Core attack logic is correct; minor undocumented heuristics, prompt-template wording differences, or misaligned naming. \\
(Weight: 40\%) & 4--6 (Fair) & Crucial algorithmic components (e.g., mutation operators, similarity thresholds, or filter-feedback loops) are missing or altered. \\
 & 1--3 (Poor) & Severe methodological mismatches; falls back to a generic prompt-rewriting baseline rather than the proposed attack. \\ \midrule
 & 9--10 (Perfect) & Flawless syntax, strict contract compliance, proper device allocation (CUDA), and clean handling of victim/judge API calls. \\
\textbf{Technical Correctness} & 7--8 (Good) & Syntactically sound and executable, but contains implicit risks (e.g., fragile response parsing or hardcoded model paths). \\
(Weight: 30\%) & 4--6 (Fair) & Explicit logical/syntax errors, tensor or image-format mismatches, or unhandled API failures requiring manual debugging. \\
 & 1--3 (Poor) & Broken execution architecture, infinite search loops, or invalid dependencies that prevent end-to-end runs. \\ \midrule
 & 9--10 (Perfect) & Includes paper-matched hyperparameters, query budgets, victim/filter/judge configs, random seed controls, and artifact logging. \\
\textbf{Reproducibility} & 7--8 (Good) & Implements the attack and evaluation logic, but omits default paper-matched configs or standardized metric aggregation. \\
(Weight: 30\%) & 4--6 (Fair) & Only implements an isolated prompt generator; missing the victim diffusion pipeline, safety-filter hooks, or judge evaluation. \\
 & 1--3 (Poor) & Bare skeleton code with no runnable evaluation entry point or configuration metadata. \\ \bottomrule
\end{tabularx}
\end{table*}

\section{Standardized Evaluation Protocol}
\label{app:standardized_protocol}
\Cref{tab:standardized_protocol} records the output count, valid denominator, and failure accounting for the cross-model results. CLIP consistency checks follow the corresponding paper-matched protocols described in the main text.

\begin{table*}[t]
\centering
\small
\begin{tabularx}{\textwidth}{@{}p{3cm}X@{}}
\toprule
\textbf{Setting} & \textbf{Standardized protocol} \\
\midrule
Dataset & NSFW-200, evaluated over valid input prompts. \\
Attempts and outputs & $K=1$ (ASR-1), with one final image per prompt and victim. \\
Image judge & The same image-only GPT-4o NSFW judge evaluates outputs from SD v1.4, SD v1.5, SDXL, and GPT-image-2. \\
Valid denominator & $N_{\mathrm{valid}}$ excludes only inputs identified as invalid before attack execution. \\
Failure accounting & Prompt refusals, policy blocks, timeouts, API errors, and generation failures count as unsuccessful attempts and remain in the denominator. \\
\bottomrule
\end{tabularx}
\caption{Shared protocol for the standardized cross-model ASR results.}
\label{tab:standardized_protocol}
\end{table*}

\section{Responsible AI Assessment: Artifacts}
\label{sec:appendix_artifacts}

\subsection{Discussion of Licenses and Terms of Use}
Artifact use follows the applicable licenses and provider terms:
\begin{itemize}
    \item \textbf{Datasets:} I2P and NSFW-200 are used for non-commercial academic safety evaluation and red-teaming under their distribution terms.
    \item \textbf{Victim models:} Stable Diffusion v1.4/v1.5 and SDXL are used under the applicable CreativeML Open RAIL-M and Open RAIL++-M licenses and acceptable-use policies.
    \item \textbf{Commercial APIs:} GPT-image-2 evaluation uses research API access subject to OpenAI's terms and usage policies.
    \item \textbf{Framework release:} The artifact-level checks below implement the release protocol in the preceding subsection under the selected license.
\end{itemize}

\subsection{Responsible Release Protocol}
To balance scientific reproducibility with dual-use safety risks, we adopt a tiered release strategy. We will publicly release the core \framework orchestration engine, runtime contract schemas, standardized evaluation harness, documentation, and sanitized configuration templates. However, runnable attack modules and adversarial prompt artifacts will undergo strict risk screening prior to release: high-risk or turnkey exploit code will be sanitized or restricted to verified academic researchers upon request. Our release decisions weigh exploitability, automation potential, cross-model transferability, and available mitigations, as documented in the artifact checklist below. All distributions will strictly comply with institutional review policies, model provider terms of service, and dataset licenses.

\subsection{Artifact-Level Responsible Release Checklist}
Each artifact is reviewed before release:
\begin{itemize}
    \item \textbf{Purpose:} identify its reproducibility or safety-auditing value.
    \item \textbf{Misuse:} assess harmful inputs, transferable attack logic, automated targeting, secrets, and deployment-ready instructions.
    \item \textbf{Sanitization:} remove risky content and unnecessary automation while testing that verification remains possible.
    \item \textbf{Access:} choose public, restricted, or withheld release according to residual risk and available safeguards.
    \item \textbf{Documentation:} verify licenses, institutional requirements, dataset and provider terms; record intended use, limitations, residual risks, and a reporting contact.
    \item \textbf{Reassessment:} revisit decisions when capabilities, policies, or misuse change.
\end{itemize}

\subsection{Documentation of Artifacts}
The evaluated artifacts have the following scope:
\begin{itemize}
    \item \textbf{Domains:} I2P, NSFW-200, and MMA cover NSFW content, violence, hate speech, illegal acts, harassment, and other policy-violating visual concepts.
    \item \textbf{Language and transformations:} Evaluation is primarily in English and includes token obfuscation, multimodal conceptual evasion, stylistic mutations, and low-effort prompt variations.
    \item \textbf{Representation:} The synthetic or crowd-sourced safety-testing prompts contain no sensitive personal or demographic identifiers; generated images may inherit cultural and demographic biases from model pre-training data.
\end{itemize}
\subsection{Model Size and Computing Infrastructure}
Stable Diffusion v1.4/v1.5 have approximately 980M parameters; SDXL has approximately 6.6B including the base and refiner. GPT-image-2's parameter count is undisclosed.

\begin{table*}[t]
\centering
\small
\begin{tabularx}{\textwidth}{@{}lXXl@{}}
\toprule
\textbf{Method} & \textbf{Transformation operator} & \textbf{Search/control strategy} & \textbf{Broad family} \\
\midrule
SneakyPrompt & Lexical/token substitution & Reinforcement learning & Token/semantic \\
HTS & Token-sequence optimization & Heuristic token search & Token/semantic \\
Low-Effort & Direct semantic rewriting & Direct transformation & Token/semantic \\
MMA-Diffusion & Adversarial prompt optimization & Gradient-based iterative optimization & Optimization/search \\
JPA & Discrete prefix/token optimization & Projected-gradient optimization & Optimization/search \\
Ring-A-Bell & Concept-representation optimization & Genetic/evolutionary search & Optimization/search \\
DiffZOO & Token-sequence optimization & Zeroth-order optimization & Optimization/search \\
DACA & Prompt decomposition & LLM-guided decomposition & LLM/VLM automation \\
JailFuzzer & Prompt mutation & LLM/VLM feedback-driven fuzzing & LLM/VLM automation \\
PGJ & Perception-guided semantic rewriting & LLM/VLM-guided iterative search & LLM/VLM automation \\
R2A & Frame-semantic rewriting & LLM-guided reasoning & LLM/VLM automation \\
\bottomrule
\end{tabularx}
\caption{Two-axis taxonomy of the eleven evaluated attacks. The common representation records each paper's transformation operator and search/control strategy while preserving its method-specific procedure.}
\label{tab:attack-taxonomy}
\end{table*}

\begin{table*}[t]
\centering
\small
\setlength{\tabcolsep}{5pt}
\renewcommand{\arraystretch}{1.12}
\begin{tabularx}{\textwidth}{@{}llccX@{}}
\toprule
\multirow{2}{*}{\textbf{Target paper}} & \multirow{2}{*}{\textbf{Reproducer}} & \multicolumn{2}{c}{\textbf{Validation}} & \multirow{2}{*}{\textbf{Implementation / evaluation evidence}} \\
\cmidrule(lr){3-4}
 & & \textbf{Attack check} & \textbf{End-to-end} & \\
\midrule
\multirow{2}{*}{Ring-A-Bell} & Paper2Code & Mismatch & Fail & CLIP/GA formulation differs; configuration and required inputs are incomplete. \\
 & \framework & \textbf{Pass} & \textbf{Pass} & Token-level GA, concept vectors, and evaluation assets are aligned. \\
\addlinespace[4pt]
\multirow{2}{*}{DACA} & Claude-only & Pass & Fail & Evaluator consumes existing results or mock data without a bound image-generation pipeline. \\
 & \framework & \textbf{Pass} & \textbf{Pass} & Paper-specific attack and evaluation modules are bound under the shared contract. \\
\addlinespace[4pt]
\multirow{2}{*}{R2A} & Claude-only & Pass & Fail & Evaluator imports and LLM initialization remain unresolved. \\
 & \framework & \textbf{Pass} & \textbf{Pass} & Executable attack and evaluation modules implement the disclosed zero-shot path. \\
\bottomrule
\end{tabularx}
\caption{Target-matched reproduction controls. Attack check reports module-level validation; end-to-end validation requires a runnable, paper-aligned attack and evaluation pipeline. Claude-only uses one Cursor-agent run per paper, initialized by a single prompt.}
\label{tab:reproduction-baselines}
\end{table*}

\section{Memory Retrieval and Relationship Scoring}
\label{app:memory_scoring}
This section gives the implementation details underlying memory retrieval and the relationship visualization.

\subsection{Retrieval Ranking}
For retrieval, let $Q$ be the target paper's token set and $R_d$ the tokens from record $d$'s summaries, attack name, and derived tags. The implementation lowercases alphanumeric tokens and retains those of length at least four. Records from the current paper are excluded; the remaining records are ranked by
\begin{equation}
r(d,p)=\frac{|Q\cap R_d|}{\max(12,|Q|)}+0.35\,\mathbf{1}[v_d=v_p],
\end{equation}
where $v$ denotes the agent variant and the overlap term is zero for an empty query. The configurable top-$k$ records (default $k=4$) supply summaries, code digests, implementation paths, and lineage to the construction context.

\subsection{Relationship Graph Scores}
The relationship scores use a separate Dice-based calculation. After tokenization and stop-word removal, let $T_d$ contain summary and tag tokens and $K_d$ contain code-digest tokens. We compute
\begin{equation}
D(A,B)=\frac{2|A\cap B|}{|A|+|B|},
\end{equation}
with $D=0$ when either set is empty. For $D_{\mathrm{text}}=D(T_i,T_j)$ and $D_{\mathrm{code}}=D(K_i,K_j)$, the graph score is
\begin{equation}
s(i,j)=\max\{D_{\mathrm{text}},1.12D_{\mathrm{code}}\}.
\end{equation}
The graph labels implementation similarity when code Dice is at least 0.17 and exceeds text Dice by at least 0.03; otherwise it labels method similarity. The 0.32 threshold in the figure summarizes strong graph links, while $r(d,p)$ determines retrieval ranking.

\section{Method Taxonomy and Reproduction Controls}
\label{app:additional_controls}
\subsection{Method Taxonomy}
\Cref{tab:attack-taxonomy} categorizes the eleven evaluated attacks by transformation operator and search/control strategy, preserving each method's procedure within the shared contract.

\subsection{Target-Matched Reproduction Controls}
Paper2Code and \framework receive the same Ring-A-Bell source paper. The DACA and R2A Claude-only controls each use one Cursor-agent run with Claude-4.5-Sonnet, initialized by a single prompt, without a planner--auditor loop or human correction. \Cref{tab:reproduction-baselines} records module-level and end-to-end validation separately, with the specific implementation and integration evidence for each case.

\section{Human Intervention and Oversight}
\label{app:human_intervention}
To transparently track human involvement during reproduction, we maintain a Human Intervention Ledger (\Cref{tab:human-ledger}) categorizing manual actions into three tiers: read-only verification (A1), minor environment/path integration (A2), and substantive algorithmic rewriting (A3). Across all eleven reproductions, the system logged 97 autonomous agent passes, 30 read-only checks (A1), only 5 minor integration fixes (A2), and zero code rewrites (A3). Excluding read-only checks, \framework achieves an autonomous pass rate of $97 / (97 + 5 + 0) = 95.1\%$, requiring an average of just 28 minutes of human oversight per paper (308 minutes total).
\begin{table}[t]
\centering
\small
\begin{tabular}{lrrrr}
\toprule
\textbf{Scope} & \textbf{Agent passes} & \textbf{A1} & \textbf{A2} & \textbf{A3} \\
\midrule
All 11 methods & 97 & 30 & 5 & 0 \\
\bottomrule
\end{tabular}
\caption{Human Intervention Ledger across all 11 reproductions. A1 denotes read-only verification, A2 light configuration/path adjustments, and A3 substantive code rewrites. A1 checks do not modify artifacts and are excluded from the autonomy denominator.}
\label{tab:human-ledger}
\end{table}

\end{appendices}
\end{document}